\documentclass[draft]{agujournal2019}
\usepackage{url} %this package should fix any errors with URLs in refs.
\usepackage{lineno}
\draftfalse

\journalname{Geophysical Research Letters}

\begin{document}
	
	\title{Offshore Wind Turbines will encounter very low Atmospheric Turbulence}
	
	\authors{Nicola Bodini\affil{1}, Julie K. Lundquist\affil{1,2}, and Anthony Kirincich\affil{3}}

	\affiliation{1}{Department of Atmospheric and Oceanic Sciences, University of Colorado Boulder, Boulder, Colorado, USA}
	\affiliation{2}{National Renewable Energy Laboratory, Golden, Colorado, USA}
	\affiliation{3}{Woods Hole Oceanographic Institution, Woods Hole, Massachusetts, USA}
	
	\correspondingauthor{Nicola Bodini}{nicola.bodini@colorado.edu}

	\begin{keypoints}
		%  Each must be 100 characters or less with no special characters or punctuation
		\item strength and persistence of wind plant wakes depends on turbulence dissipation rate
		\item turbulence dissipation rate offshore is small with a weak diurnal cycle
		\item dissipation rate is larger when flow is from the land, which tends to be in wintertime at this site
	\end{keypoints}

	\begin{abstract}
		The rapid growth of offshore wind energy requires accurate modeling of the wind resource, which can be depleted by wind farm wakes. Turbulence dissipation rate ($\epsilon$) governs the accuracy of model predictions of hub-height wind speed and the development and erosion of wakes. Here we assess the variability of turbulence kinetic energy and $\epsilon$ using 13 months of observations from a profiling lidar deployed on a platform off the Massachusetts coast.  Offshore, $\epsilon$ is 2 orders of magnitude smaller than onshore, with a subtle diurnal cycle. Wind direction largely influences the annual cycle of turbulence, with larger values in winter when the wind flows from the land, and smaller values in summer, when the wind is mainly from open ocean. Because of the weak turbulence, wind plant wakes will be stronger and persist farther downwind in summer.
	\end{abstract}

	\section{Introduction}
	
	Wind energy continues to expand as one of the cleanest energy technologies, with its zero carbon emissions \cite{boyle2004renewable} and zero operational water consumption \cite{macknick2012operational}.
	The faster and steadier winds which blow over the low-friction surface of the ocean \cite{landberg2015meteorology} represent a valuable source of clean energy, especially given that the cost of offshore wind energy is decreasing faster than expected \cite{stiesdal_2016}. Many favorable locations for offshore wind farm development are close to coastal areas with large energy needs \cite{manwell2010wind}, minimizing the need for long-distance transmission \cite{wiser2015wind}.\\
	Currently, most of the existing offshore wind farms are located in Northern Europe, where they account for a capacity of about 15 GW, with a planned increase to about 74 GW by 2030 \cite{hoof2017unlocking}.
	In the United States, only a single 30 MW commercial offshore wind farm \cite{wind2016block} has been built. However, the U.S. offshore technical resource potential is estimated to be nearly double the nation's current electricity use \cite{musial2016offshore}. Many offshore wind projects are currently being planned, mostly concentrated along the Eastern Seaboard. The State of Massachusetts plans to procure 1,600 MW of installed offshore wind, representing about 11\% of its overall needs, by 2027 \cite{musial20172016}, with beneficial impacts on the State's economy and employment \cite{center20182018}.\\
	
	As offshore wind energy development grows, accurate forecasting of the available wind resource in the offshore environment is required.
	Recent studies \cite{yang2017sensitivity,berg2018sensitivity} have shown that the hub-height wind speed predicted by the Weather Research and Forecasting (WRF) model \cite{skamarock2005description} is highly sensitive to the parametrization of turbulence dissipation rate ($\epsilon$), which is responsible for up to 50\% of the variance in hub-height wind speed. Current parametrizations of $\epsilon$ assume a local balance between production and dissipation of turbulence within a grid cell. However, this assumption does not hold when using models at a fine horizontal resolution \cite{nakanishi2006improved,krishnamurthy2011wind, hong2012next}. Therefore, improved representations of $\epsilon$ in models are crucially needed to enhance the accuracy of wind energy forecasting.\\
	Turbulence also plays a key role in the development and subsequent erosion of wind turbine and wind farm wakes, whose spatial extent has been observed to be particularly long offshore \cite{platis2018first,siedersleben2018evaluation,siedersleben2018micrometeorological}, with observed wakes extending beyond 45 km in some cases. An accurate model representation of the turbulence dissipation rate would allow for a better layout optimization of offshore wind farms, which would in turn reduce the large costs related to wind farm wakes \cite{nygaard2014wakes} deriving from the uncoordinated development of wind projects \cite{lundquist2019costs}. Therefore, an assessment of the variability of $\epsilon$ from observations in the offshore atmospheric boundary layer is an essential first step towards reducing uncertainty in offshore wind energy forecasting and optimizing energy production.\\
	
	Various techniques to retrieve turbulence dissipation rates from sonic anemometers \cite{champagne1977flux,oncley1996surface}, high-frequency hot-wire anemometers suspended on tethered lifting systems \cite{frehlich2006measurements,lundquist2015dissipation} or flown on aircrafts \cite{fairall1980aircraft} or UAVs \cite{lawrence2013high} have been developed. The ease of deployment and extended measurement range of remote sensing instruments have also fueled research to derive methods to retrieve $\epsilon$ from lidars \cite{frehlich1994coherent,banakh1996measurement,o2010method,wulfmeyer2016determination} and radars \cite{shaw2003turbulence,mccaffrey2017improved}. The extension and application of these techniques has led to a systematic assessment of the variability of $\epsilon$ in both flat \cite{bodini2018estimation} and complex terrain \cite{bodini2019variability}. However, to the authors' knowledge, no comprehensive analysis of the variability of turbulence dissipation rate in offshore environment has been completed.\\
	
	Here, we assess the temporal variability of turbulence dissipation rate retrieved from 13 months of observations from a wind-profiling lidar deployed on an offshore platform. In Section 2 we describe the site off the Massachusetts coast, and the method used to retrieve $\epsilon$. Section 3 presents the daily and seasonal variability of $\epsilon$ and its relation to wind direction and land influence. In Section 4 we summarize and conclude our analysis.

	\section{Data and Methods}
	
	\subsection{The Massachusetts MetOcean Initiative}
\begin{figure}[h]
		\centering
		\includegraphics[width=28pc]{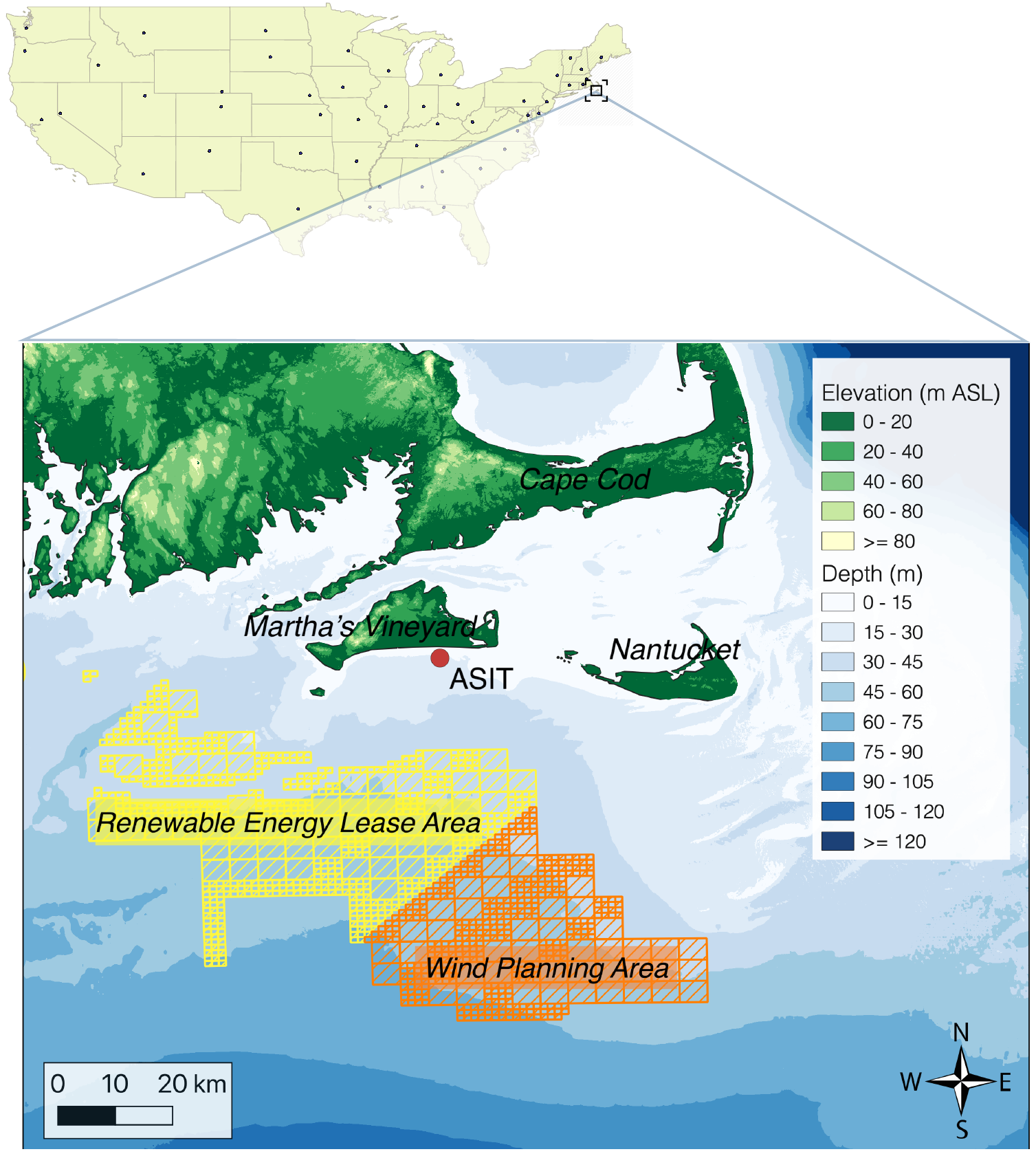}
		\caption{Map of the areas off the coasts of Rhode Island and Massachusetts where future development of offshore wind projects will take place. The location of the ASIT platform, where the WINDCUBE v2 lidar used in this study was deployed, is shown.}
		\label{Fig1}
	\end{figure}

	A "MetOcean Initiative" \cite{filippelli2015meocean} funded by the Massachusetts Clean Energy 
	Center has collected continuous observations of the atmospheric boundary layer at the Woods Hole Oceanographic Institution's (WHOI) Air-Sea Interaction Tower (ASIT) since October 2016.  The ASIT is a cabled, fixed platform located approximately 3 km south of Martha's Vineyard in 17 m of water (Figure \ref{Fig1}) and proximate to the Rhode Island and Massachusetts Wind Energy Areas, which represent the US's largest region under development for offshore wind energy extraction.
	At the site, a suite of wind resource monitoring equipment was used to augment the existing sensors deployed by WHOI's Martha's Vineyard Coastal Observatory (MVCO), including a pair of cup anemometers above the top of the tower at 26 m above mean sea level (ASL), a wind vane at 23 m ASL, and a WINDUCUBE version (v2) profiling lidar on the main platform, at 13 m ASL.
	All metocean data collected by WHOI for the project was validated by AWS Truepower.
	The v2 lidar measures line-of-sight velocity along the 4 cardinal directions with a nominal zenith angle of 28$^{\circ}$, with an additional line-of-sight velocity measurement along the vertical. 
	The lidar was positioned on a platform extending away from the ASIT mast into the direction of prevailing winds (to the southwest) and rotated such that none of the measurement beams of the lidar were affected by the tower structure. Approximately 5 s are needed to complete the five-beam cycle. The lidar took measurements at 10 heights: 53, 60, 80, 90, 100, 120, 140, 160, 180, and 200 meters above sea level.  Thus, the lidar-measured winds were above the level of, and not affected by the tower structure, nor were they expected to exhibit any signal of wave motion.\\
	Here, the first 13 months of observations, from 7 October 2016 through 29 October 2017, have been analyzed. Periods during which precipitation was recorded at WHOI's shore lab on the southern coast of Martha's Vineyard have been excluded from the analysis.

	\subsection{Turbulence dissipation rate from profiling lidars}
	
	Turbulence dissipation rate can be estimated from the variance of the line-of-sight velocity measured by profiling lidars following the approach described in \citeA{o2010method} and refined in \citeA{bodini2018estimation}, with the assumption of locally homogeneous and isotropic turbulence.
	This approach derives $\epsilon$ by integrating the turbulence spectrum within the inertial subrange. To do so, the maximum length scale (and thus the sample size) to include in the calculation must be accurately chosen \cite{tonttila2015turbulent,bodini2018estimation}. Here we use a local regression of the spectrum of the line-of-sight velocity to estimate the extension of the inertial subrange, as described and tested in \citeA{bodini2019variability}. The distribution of sample size values we obtain is between 20 s (5th percentile) and 300 s (95th percentile). $\epsilon$ is then calculated as
	\begin{linenomath*}
		\begin{equation}
		\epsilon = 2 \pi \left( \frac{2}{3a} \right)^{3/2} \left( \frac{\sigma_b^2}{L_N^{2/3}-L_1^{2/3}} \right)^{3/2},
		\label{epslidar}
		\end{equation}
	\end{linenomath*}
	where $a =$ 0.52 is the one-dimensional Kolmogorov constant \cite{paquin1971determination,sreenivasan1995universality}, $L_1 = U t$, with $U$ the horizontal wind speed and $t$ the dwell time, and $L_N = N L_1$, where $N$ is the size of the line-of-sight velocity sample determined from the local regression of the experimental spectra. The variance $\sigma_b^2$ is calculated by subtracting a contribution due to the instrumental noise from the variance (averaged across the lidar beams) of line-of-sight velocity $\sigma_v^2$: $\sigma_b^2 = \sigma_v^2-\sigma_e^2$, where $\sigma_e^2$ is defined as in equation 2 in \citeA{pearson2009analysis}. More details of this method are available in \citeA{bodini2018estimation,bodini2019variability}.

	\section{Results}
	
\begin{figure}[h]
		\centering
		\includegraphics[width=0.65\textwidth]{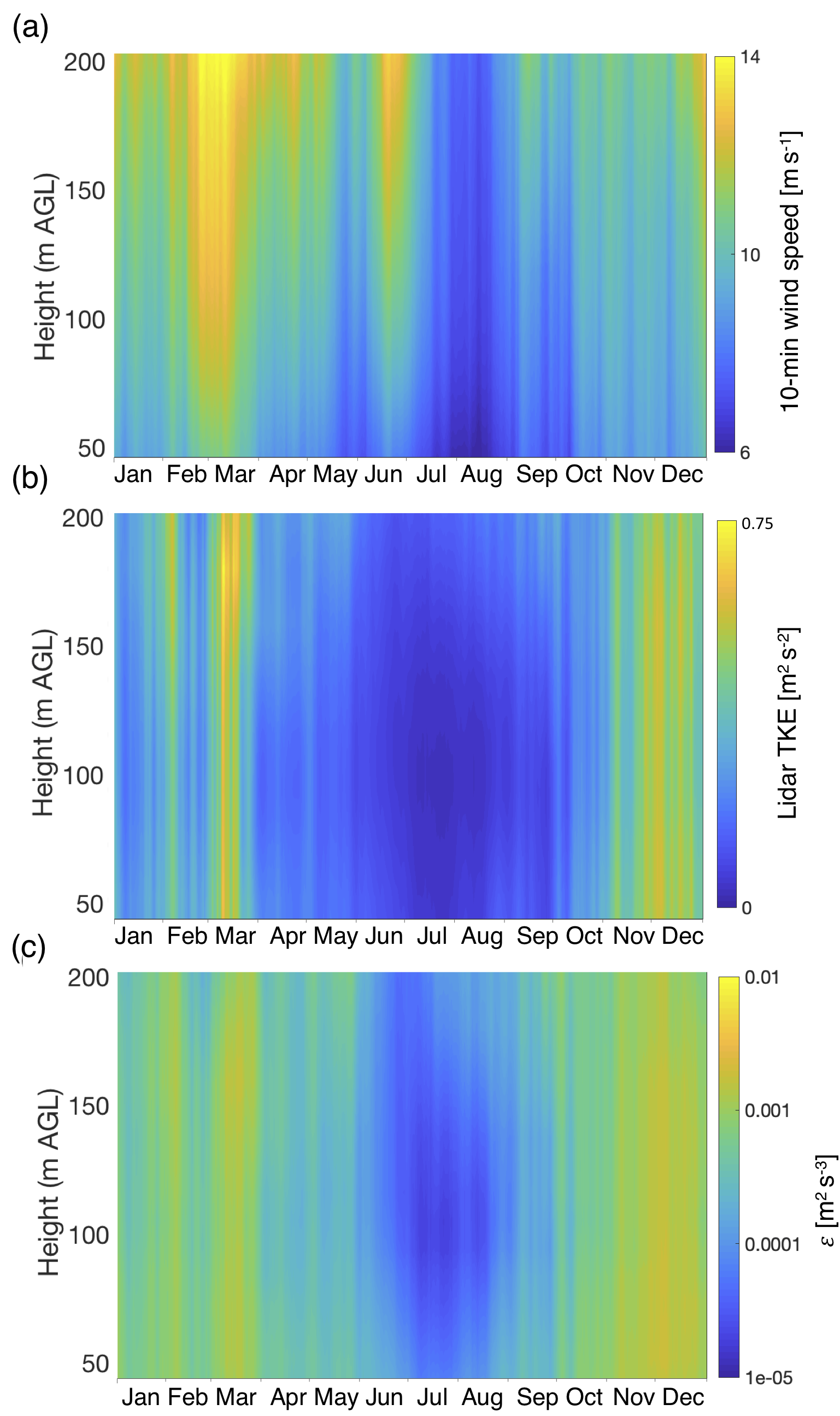}
		\caption{Annual cycle of (a) wind speed, (b) lidar TKE, and (c) turbulence dissipation rate. At each height, data have been smoothed using a 30-day running mean.}
		\label{Fig2}
	\end{figure}

	Turbulence dissipation occurs in an environment determined by wind speed, whose annual cycle at the site is shown in Figure \ref{Fig2}a (data have been smoothed with a 30-day running mean). 
	Wind speed is generally strong during winter, with average values above 10 m s$^{-1}$ at wind-turbine-hub-heights throughout the season, while summer shows smaller values, with minima in August, when the average wind speeds were less than 10 m s$^{-1}$ at every observed height. Turbulence quantities also exhibit an annual cycle. We calculate turbulence kinetic energy (TKE) as
	\begin{linenomath*}
		\begin{equation}
		TKE = \frac{1}{2} \left( \sigma_u^2 + \sigma_v^2 + \sigma_w^2 \right)
		\end{equation}
	\end{linenomath*}
	where the variances of the wind components are calculated over 2-min intervals. As noted by \citeA{sathe2011can}, lidars cannot fully resolve the wind variances, as a sonic anemometer would, given the lidars' limited temporal frequency. However, since data from a sonic anemometer are not available in this case, we calculate TKE using data from the lidar, and will refer to it as lidar TKE \cite{rhodes_effect_2013,kumer2016turbulent}. The annual cycle of lidar TKE (Figure \ref{Fig2}b) reveals a clear pattern, with extremely small values during the summer, and larger values in winter, with little dependence on height.
	The annual variability of turbulence dissipation (Figure \ref{Fig2}c) follows a similar pattern, and reaches maximum in fall and winter, with minima in the summer. Monthly average values of $\epsilon$ have a large correlation with monthly average lidar TKE ($R = 0.88$): when the kinetic energy of turbulence is on average large, large values of $\epsilon$ are usually needed to dissipate this energy. Monthly average wind speed has a slightly smaller correlation with $\epsilon$ ($R = 0.71$), with some considerable discrepancies (e.g. June).
	The annual cycle of $\epsilon$ at this offshore location differs from what was measured in similar campaigns onshore, where the annual cycle of $\epsilon$ is mainly driven by the seasonal cycle of convection, with larger turbulence in summer due to increased convection, and more quiescent conditions in winter due to higher stratification \cite{bodini2019variability}.\\
	
\begin{figure}[h]
		\centering
		\includegraphics[width=1\textwidth]{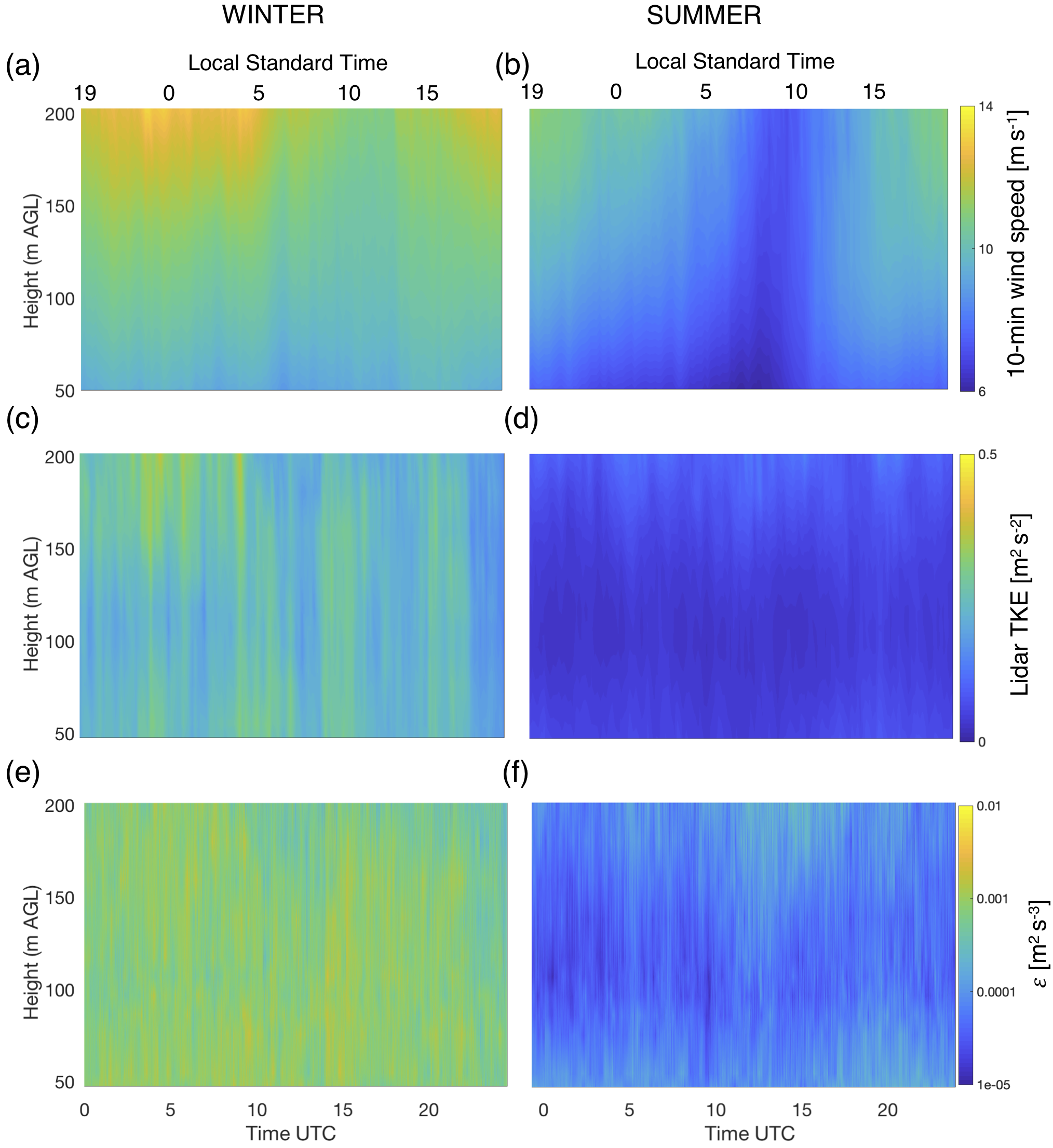}
		\caption{Average diurnal climatologies of wind speed, lidar TKE and turbulence dissipation rate for December, January and February (panels a, c, e) and June, July and August (panels b, d, f).}
		\label{Fig3}
	\end{figure}

	The annual variability of wind speed and turbulent properties at the site has a considerable impact on the average diurnal climatologies of these variables in different seasons (Figure \ref{Fig3}).
	Wind speed tends to increase during the late afternoon and at night, while in daytime conditions lower average wind speeds occur, with weaker shear, as also found by \citeA{archer2016predominance}. On average, summer months (Figure \ref{Fig3}b) show a larger diurnal cycle of wind speed compared to winter (Figure \ref{Fig3}a). 
	In contrast to this diurnal cycle in wind speed, the diurnal cycle of lidar TKE (Figure \ref{Fig3}c and d) is subtle, with a small variability with height. The average diurnal cycles of $\epsilon$ offshore (Figure \ref{Fig3}e and f) show similar patterns to lidar TKE, with differences greater than one order of magnitude between winter and summer. In summer, $\epsilon$ shows local minima at $\sim$100 m ASL, and increased values at $\sim$200 m ASL, especially in the local morning. In contrast, in winter $\epsilon$ shows a minimum in the morning at $\sim$200 m ASL. The average differences throughout the day in $\epsilon$ are smaller than one order of magnitude, whereas the diurnal cycle of $\epsilon$ onshore, in both flat \cite{bodini2018estimation} and complex \cite{bodini2019variability} terrain, shows a larger amplitude, with differences of at least one order of magnitude between larger values during daytime and smaller values in nighttime quiescent conditions.	Moreover, the average values of $\epsilon$ are smaller (in some cases of over an order of magnitude) than onshore \cite{bodini2018estimation,bodini2019variability}.\\
	
\begin{figure}[h]
		\centering
		\includegraphics[width=1\textwidth]{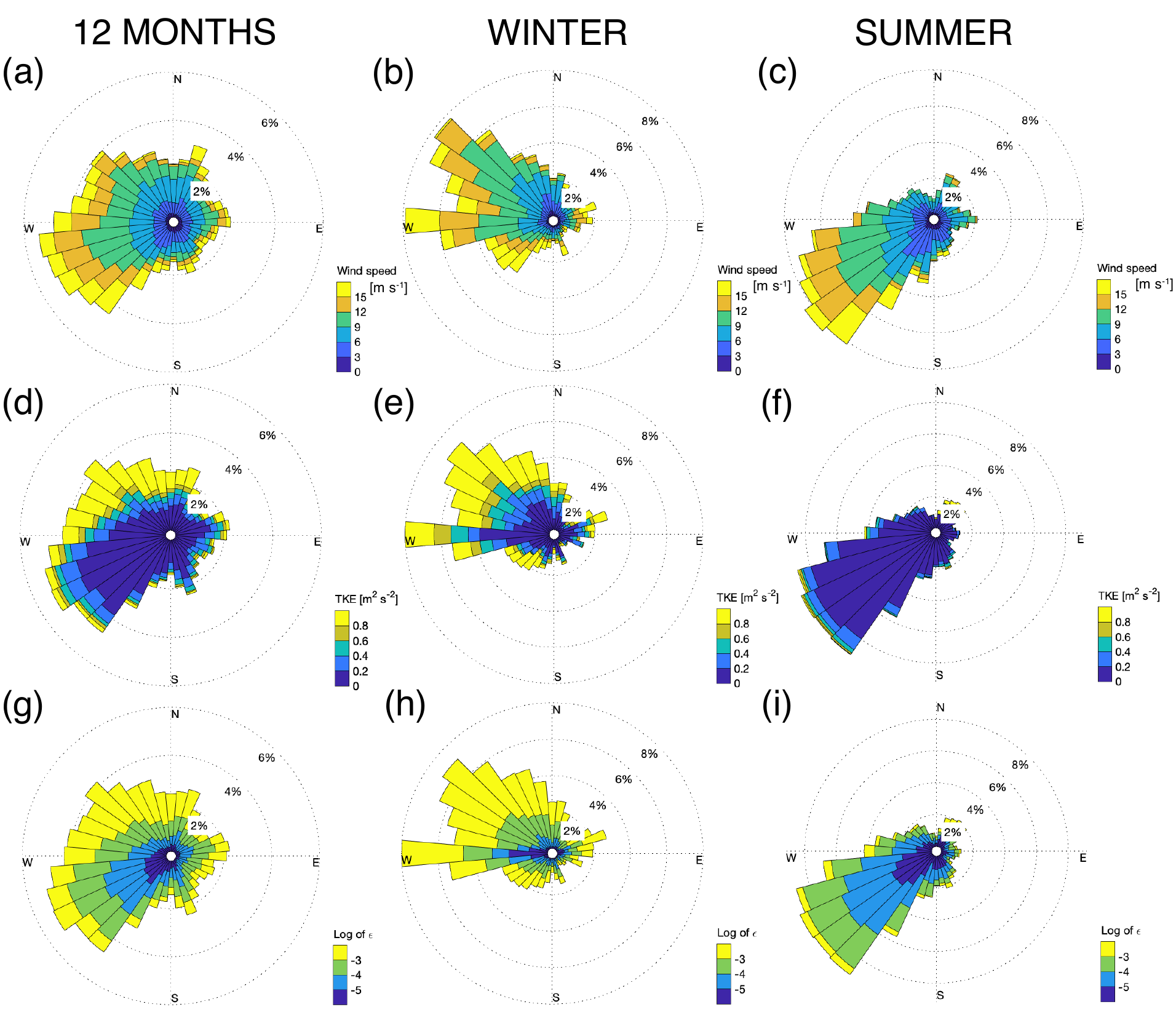}
		\caption{Roses of (a-c) wind speed, (d-f) lidar TKE and (g-i) turbulence dissipation rate, at 100 m, from: the (a, d, g) full 13-month period considered; (b, e, h) December, January and February; and (c, f, i) June, July and August.}
		\label{Fig4}
	\end{figure}

	The summer minimum in lidar TKE and turbulence dissipation rate occurs because of wind regimes. North-westerly, westerly and south-westerly winds dominate at the ASIT site (Figure \ref{Fig4}a). The wind direction dictates turbulence properties: north-westerly winds generally lead to large values of lidar TKE and $\epsilon$. In contrast, south-westerly winds generally cause low turbulence regimes (panels d, g).	
	This relationship between wind direction and turbulence can be explained by considering the location of the ASIT platform (Figure \ref{Fig1}). When the wind comes from the north-west, the flow interacts with the land before reaching the offshore platform. This land wake effect generates turbulence, both in terms of lidar TKE and $\epsilon$. On the contrary, southwesterly winds come from the open ocean, where lower friction causes smaller values of lidar TKE and turbulence dissipation rates. In the summer (June, July, August), the winds consist of almost exclusively south-westerly winds (Figure \ref{Fig4}c), associated with lidar TKE generally smaller than 0.2 m$^2$ s$^{-2}$ (at 100 m ASL, Figure \ref{Fig4}f) and turbulence dissipation rarely exceeding 10$^{-3}$ m$^2$ s$^{-3}$ (Figure \ref{Fig4}i). On the contrary, during the winter, northwesterly winds dominate (Figure \ref{Fig4}b), with larger TKE (Figure \ref{Fig4}c) and $\epsilon$ (Figure \ref{Fig4}h). The annual cycle of turbulence dissipation rate offshore is more influenced by the wind-land interaction rather than the seasonal cycle itself.\\
	
\begin{figure}[h]
		\centering
		\includegraphics[width=0.55\textwidth]{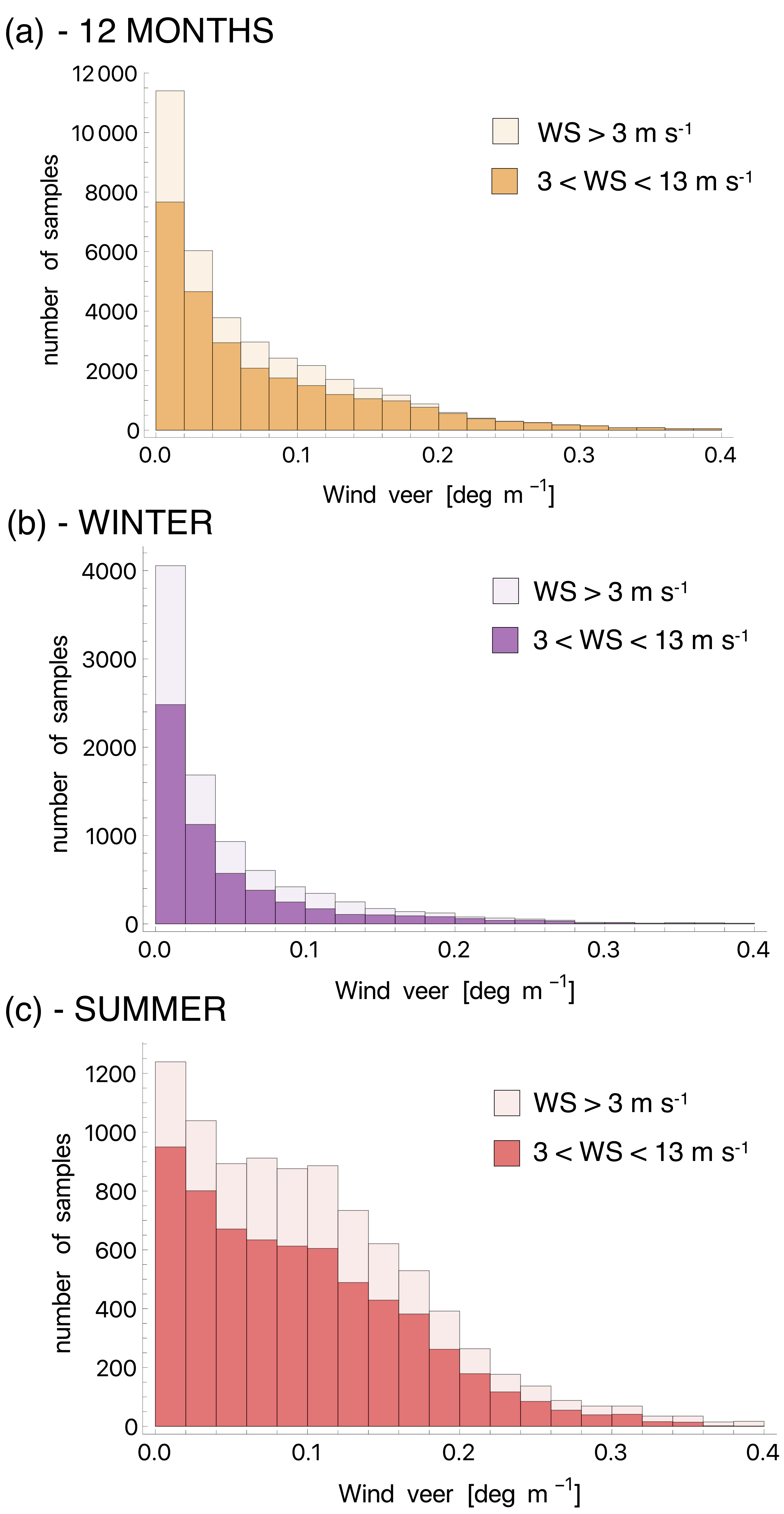}
		\caption{Histograms of 2-minute average wind veer between 40 m and 200 m ASL from the 13 months of observations (panel a), June, July, August (panel b), and December, January, February (panel c).}
		\label{Fig5}
	\end{figure}

	An annual cycle also emerges in wind veer, another important atmospheric variable which affects the structure of wind turbine wakes \cite{bodini2017three,abkar2018analytical,churchfield2018effects}. We calculate wind veer as the difference in 2-minute average wind direction, retrieved from the lidar, between 40 m and 200 m ASL, which represent typical vertical limits for the rotor of modern offshore wind turbine models. We present the observations as change in wind direction per meter of height ($^{\circ}$ m$^{-1}$) to normalize the results in terms of the vertical separation between the measurements. Histograms of wind veer are shown in Figure \ref{Fig5}, where we also highlight wind veer values which were measured with an average wind speed between 3 m s$^{-1}$ and 13 m s$^{-1}$, which correspond to region 2, the area of the power curve where power is more sensitive to wind speed \cite{manwell2010wind}, of the Siemens Gamesa 7.0 MW turbine.
	Wind veer often assumes large values between the considered vertical limits, with an average value of 0.07$^{\circ}$ m$^{-1}$ (0.08$^{\circ}$ m$^{-1}$ for region 2 wind speeds) when considering the whole period of observations (panel a). However, important differences between the seasons emerge. In winter months (panel b), the average wind veer wind observed (0.05$^{\circ}$ m$^{-1}$, 0.04$^{\circ}$ m$^{-1}$ for regime 2 wind speeds) is only half of the summer average (0.10$^{\circ}$ m$^{-1}$, 0.09$^{\circ}$ m$^{-1}$ for regime 2 wind speeds). The summertime offshore wind veer conditions are similar to the nighttime stable conditions found onshore in \citeA{walter2009speed}. Veer is of particular interest with respect to wind turbine wake propagation \cite{bodini2017three,churchfield2018effects}, and wakes impact power production the most at wind speeds in region 2. The coupling of the strong veer in summertime conditions with low dissipation will result in long-propagating but skewed wakes, impacting power production and turbulent loads on downwind turbines.

	\section{Discussion and Conclusions}
	
	Offshore wind plant wakes can extend tens of km downwind \cite{platis2018first,siedersleben2018evaluation,siedersleben2018micrometeorological} in low-turbulence, stably-stratified conditions. These wakes undermine offshore power production \cite{nygaard2014wakes}. Because wind plant wake propagation is influenced by turbulence variability \cite{lundquist2019costs}, we assess the turbulence dissipation rate ($\epsilon$) from a year-long dataset of offshore wind lidar deployment.\\
	We have retrieved $\epsilon$ from 13 months of observations from a profiling lidar located on an offshore platform $\sim$ 3 km south of Martha's Vineyard, off the coast of Massachusetts. Offshore $\epsilon$ has, on average, smaller values compared to those found in comparable studies onshore, with a weak diurnal cycle. The small average values of $\epsilon$ are conducive to strong and long-lived wind plant wakes such as those wakes observed in stable stratification in the North Sea \cite{platis2018first,siedersleben2018evaluation,siedersleben2018micrometeorological}. Moreover, this extended persistence of wakes will be noticeable throughout the diurnal cycle, given the absence of strong turbulence dissipation even in convective daytime conditions. The seasonal cycle of $\epsilon$ is largely influenced by the dominant wind directions at the site. When the wind comes from the land, the interaction between the wind flow and the onshore topography increases the TKE (which has a large positive correlation with $\epsilon$) observed offshore and, consequently, turbulence dissipation. Therefore, the study of the optimal layout of offshore wind farms needs to consider the different turbulence regimes associated with the dominant wind patterns at each site, to possibly take advantage of the faster wake erosion connected to larger dissipation values caused by land wake effects.\\
	While the offshore wind resource is considerable, and the daily timing of wind speed increases due to sea breeze effects in the summer are particularly valuable for integrating wind energy into power grids, these results suggest that flow from the south and southwest that would lead to increased wind speeds would also have reduced turbulence, leading to stronger and more persistent wind plant wakes. Moreover, wakes in the summer would be affected by wind veer greater than 14$^{\circ}$ between the vertical limits of typical current commercial offshore wind turbines, and even larger values can be expected when considering wind gusts \cite{worsnop2017gusts}. This large wind veer can impact the effectiveness of wake steering solutions \cite{fleming2017field,fleming2019initial} to minimize the wake energy loss.\\
	Moreover, as the current wind farm lease areas are more than 25 km offshore of the platform where the profiling lidar was deployed, even the stronger turbulence conditions observed in the winter could represent an extreme upper bound on the boundary layer turbulence in the lease areas. The increased turbulence produced by the flow interaction with the land would likely dissipate as it flows offshore. As offshore wind plants in this region are developed, consideration of the likely persistence of wind plant wakes will be required to accurately predict the wind resource as well as the effect of skewed wakes on turbine operations and maintenance costs.\\
	Given the complexity of the dependencies between $\epsilon$ and other atmospheric variables, as well as the importance of the interaction with the site-specific topography, the potential of sophisticated machine learning techniques could be tested to improve the model parametrizations of $\epsilon$, as already successfully done with other atmospheric phenomena \cite{sharma2011predicting,xingjian2015convolutional,alemany2018predicting,gentine2018could}.

	\acknowledgments
	Collection of the underlying wind data that provides the basis for this analysis was funded by the Massachusetts Clean Energy Center through agreements with the Woods Hole Oceanographic Institution and AWS Truepower. The authors appreciate the efforts of the MVCO/ASIT marine technicians at WHOI and AWS staff who helped collect the data. This analysis was supported by the National Science Foundation CAREER Award (AGS-1554055) to JKL and NB, and by internal funds from the Woods Hole Oceanographic Institution for AK. The lidar observations used here are available by contacting Dr. Kirincich directly (akirincich@whoi.edu).
	\bibliography{nbbiblio}

\begin{thebibliography}{}

\bibitem [\protect \citeauthoryear {%
Abkar%
, S{\o}rensen%
\BCBL {}\ \BBA {} Port{\'e}-Agel%
}{%
Abkar%
\ \protect \BOthers {.}}{%
{\protect \APACyear {2018}}%
}]{%
abkar2018analytical}
\APACinsertmetastar {%
abkar2018analytical}%
\begin{APACrefauthors}%
Abkar, M.%
, S{\o}rensen, J.%
\BCBL {}\ \BBA {} Port{\'e}-Agel, F.%
\end{APACrefauthors}%
\unskip\
\newblock
\APACrefYearMonthDay{2018}{}{}.
\newblock
{\BBOQ}\APACrefatitle {An Analytical Model for the Effect of Vertical Wind Veer
  on Wind Turbine Wakes} {An analytical model for the effect of vertical wind
  veer on wind turbine wakes}.{\BBCQ}
\newblock
\APACjournalVolNumPages{Energies}{11}{7}{1838}.
\PrintBackRefs{\CurrentBib}

\bibitem [\protect \citeauthoryear {%
Alemany%
, Beltran%
, Perez%
\BCBL {}\ \BBA {} Ganzfried%
}{%
Alemany%
\ \protect \BOthers {.}}{%
{\protect \APACyear {2018}}%
}]{%
alemany2018predicting}
\APACinsertmetastar {%
alemany2018predicting}%
\begin{APACrefauthors}%
Alemany, S.%
, Beltran, J.%
, Perez, A.%
\BCBL {}\ \BBA {} Ganzfried, S.%
\end{APACrefauthors}%
\unskip\
\newblock
\APACrefYearMonthDay{2018}{}{}.
\newblock
{\BBOQ}\APACrefatitle {Predicting Hurricane Trajectories using a Recurrent
  Neural Network} {Predicting hurricane trajectories using a recurrent neural
  network}.{\BBCQ}
\newblock
\APACjournalVolNumPages{arXiv preprint arXiv:1802.02548}{}{}{}.
\PrintBackRefs{\CurrentBib}

\bibitem [\protect \citeauthoryear {%
Archer%
, Colle%
, Veron%
, Veron%
\BCBL {}\ \BBA {} Sienkiewicz%
}{%
Archer%
\ \protect \BOthers {.}}{%
{\protect \APACyear {2016}}%
}]{%
archer2016predominance}
\APACinsertmetastar {%
archer2016predominance}%
\begin{APACrefauthors}%
Archer, C\BPBI L.%
, Colle, B\BPBI A.%
, Veron, D\BPBI L.%
, Veron, F.%
\BCBL {}\ \BBA {} Sienkiewicz, M\BPBI J.%
\end{APACrefauthors}%
\unskip\
\newblock
\APACrefYearMonthDay{2016}{}{}.
\newblock
{\BBOQ}\APACrefatitle {On the predominance of unstable atmospheric conditions
  in the marine boundary layer offshore of the US northeastern coast} {On the
  predominance of unstable atmospheric conditions in the marine boundary layer
  offshore of the us northeastern coast}.{\BBCQ}
\newblock
\APACjournalVolNumPages{Journal of Geophysical Research:
  Atmospheres}{121}{15}{8869--8885}.
\PrintBackRefs{\CurrentBib}

\bibitem [\protect \citeauthoryear {%
Banakh%
, Werner%
, K{\"o}pp%
\BCBL {}\ \BBA {} Smalikho%
}{%
Banakh%
\ \protect \BOthers {.}}{%
{\protect \APACyear {1996}}%
}]{%
banakh1996measurement}
\APACinsertmetastar {%
banakh1996measurement}%
\begin{APACrefauthors}%
Banakh, V.%
, Werner, C.%
, K{\"o}pp, F.%
\BCBL {}\ \BBA {} Smalikho, I.%
\end{APACrefauthors}%
\unskip\
\newblock
\APACrefYearMonthDay{1996}{}{}.
\newblock
{\BBOQ}\APACrefatitle {Measurement of the turbulent energy dissipation rate
  with a scanning Doppler lidar} {Measurement of the turbulent energy
  dissipation rate with a scanning doppler lidar}.{\BBCQ}
\newblock
\APACjournalVolNumPages{ATMOSPHERIC AND OCEANIC OPTICS C/C OF OPTIKA ATMOSFERY
  I OKEANA}{9}{}{849--853}.
\PrintBackRefs{\CurrentBib}

\bibitem [\protect \citeauthoryear {%
Berg%
\ \protect \BOthers {.}}{%
Berg%
\ \protect \BOthers {.}}{%
{\protect \APACyear {2018}}%
}]{%
berg2018sensitivity}
\APACinsertmetastar {%
berg2018sensitivity}%
\begin{APACrefauthors}%
Berg, L\BPBI K.%
, Liu, Y.%
, Yang, B.%
, Qian, Y.%
, Olson, J.%
, Pekour, M.%
\BDBL {}Hou, Z.%
\end{APACrefauthors}%
\unskip\
\newblock
\APACrefYearMonthDay{2018}{}{}.
\newblock
{\BBOQ}\APACrefatitle {Sensitivity of Turbine-Height Wind Speeds to Parameters
  in the Planetary Boundary-Layer Parametrization Used in the Weather Research
  and Forecasting Model: Extension to Wintertime Conditions} {Sensitivity of
  turbine-height wind speeds to parameters in the planetary boundary-layer
  parametrization used in the weather research and forecasting model: Extension
  to wintertime conditions}.{\BBCQ}
\newblock
\APACjournalVolNumPages{Boundary-Layer Meteorology}{}{}{1--12}.
\PrintBackRefs{\CurrentBib}

\bibitem [\protect \citeauthoryear {%
Bodini%
, Lundquist%
, Krishnamurthy%
, Pekour%
\BCBL {}\ \BBA {} Berg%
}{%
Bodini%
\ \protect \BOthers {.}}{%
{\protect \APACyear {2019}}%
}]{%
bodini2019variability}
\APACinsertmetastar {%
bodini2019variability}%
\begin{APACrefauthors}%
Bodini, N.%
, Lundquist, J\BPBI K.%
, Krishnamurthy, R.%
, Pekour, M.%
\BCBL {}\ \BBA {} Berg, L\BPBI K.%
\end{APACrefauthors}%
\unskip\
\newblock
\APACrefYearMonthDay{2019}{}{}.
\newblock
{\BBOQ}\APACrefatitle {Spatial and temporal variability of turbulence
  dissipation rate in complex terrain} {Spatial and temporal variability of
  turbulence dissipation rate in complex terrain}.{\BBCQ}
\newblock
\APACjournalVolNumPages{Atmospheric Chemistry and Physics Discussions}{in
  review}{-}{-}.
\PrintBackRefs{\CurrentBib}

\bibitem [\protect \citeauthoryear {%
Bodini%
, Lundquist%
\BCBL {}\ \BBA {} Newsom%
}{%
Bodini%
\ \protect \BOthers {.}}{%
{\protect \APACyear {2018}}%
}]{%
bodini2018estimation}
\APACinsertmetastar {%
bodini2018estimation}%
\begin{APACrefauthors}%
Bodini, N.%
, Lundquist, J\BPBI K.%
\BCBL {}\ \BBA {} Newsom, R\BPBI K.%
\end{APACrefauthors}%
\unskip\
\newblock
\APACrefYearMonthDay{2018}{}{}.
\newblock
{\BBOQ}\APACrefatitle {Estimation of turbulence dissipation rate and its
  variability from sonic anemometer and wind Doppler lidar during the {XPIA}
  field campaign} {Estimation of turbulence dissipation rate and its
  variability from sonic anemometer and wind doppler lidar during the {XPIA}
  field campaign}.{\BBCQ}
\newblock
\APACjournalVolNumPages{Atmospheric Measurement Techniques}{11}{7}{4291--4308}.
\PrintBackRefs{\CurrentBib}

\bibitem [\protect \citeauthoryear {%
Bodini%
, Zardi%
\BCBL {}\ \BBA {} Lundquist%
}{%
Bodini%
\ \protect \BOthers {.}}{%
{\protect \APACyear {2017}}%
}]{%
bodini2017three}
\APACinsertmetastar {%
bodini2017three}%
\begin{APACrefauthors}%
Bodini, N.%
, Zardi, D.%
\BCBL {}\ \BBA {} Lundquist, J\BPBI K.%
\end{APACrefauthors}%
\unskip\
\newblock
\APACrefYearMonthDay{2017}{}{}.
\newblock
{\BBOQ}\APACrefatitle {Three-dimensional structure of wind turbine wakes as
  measured by scanning lidar} {Three-dimensional structure of wind turbine
  wakes as measured by scanning lidar}.{\BBCQ}
\newblock
\APACjournalVolNumPages{Atmospheric Measurement Techniques}{10}{8}{}.
\PrintBackRefs{\CurrentBib}

\bibitem [\protect \citeauthoryear {%
Boyle%
}{%
Boyle%
}{%
{\protect \APACyear {2004}}%
}]{%
boyle2004renewable}
\APACinsertmetastar {%
boyle2004renewable}%
\begin{APACrefauthors}%
Boyle, G.%
\end{APACrefauthors}%
\unskip\
\newblock
\APACrefYear{2004}.
\newblock
\APACrefbtitle {Renewable energy} {Renewable energy}.
\newblock
\APACaddressPublisher{}{Oxford University Press}.
\PrintBackRefs{\CurrentBib}

\bibitem [\protect \citeauthoryear {%
Champagne%
, Friehe%
, LaRue%
\BCBL {}\ \BBA {} Wynagaard%
}{%
Champagne%
\ \protect \BOthers {.}}{%
{\protect \APACyear {1977}}%
}]{%
champagne1977flux}
\APACinsertmetastar {%
champagne1977flux}%
\begin{APACrefauthors}%
Champagne, F.%
, Friehe, C.%
, LaRue, J.%
\BCBL {}\ \BBA {} Wynagaard, J.%
\end{APACrefauthors}%
\unskip\
\newblock
\APACrefYearMonthDay{1977}{}{}.
\newblock
{\BBOQ}\APACrefatitle {Flux measurements, flux estimation techniques, and
  fine-scale turbulence measurements in the unstable surface layer over land}
  {Flux measurements, flux estimation techniques, and fine-scale turbulence
  measurements in the unstable surface layer over land}.{\BBCQ}
\newblock
\APACjournalVolNumPages{Journal of the Atmospheric Sciences}{34}{3}{515--530}.
\PrintBackRefs{\CurrentBib}

\bibitem [\protect \citeauthoryear {%
Churchfield%
\ \BBA {} Sirnivas%
}{%
Churchfield%
\ \BBA {} Sirnivas%
}{%
{\protect \APACyear {2018}}%
}]{%
churchfield2018effects}
\APACinsertmetastar {%
churchfield2018effects}%
\begin{APACrefauthors}%
Churchfield, M\BPBI J.%
\BCBT {}\ \BBA {} Sirnivas, S.%
\end{APACrefauthors}%
\unskip\
\newblock
\APACrefYearMonthDay{2018}{}{}.
\newblock
{\BBOQ}\APACrefatitle {On the Effects of Wind Turbine Wake Skew Caused by Wind
  Veer} {On the effects of wind turbine wake skew caused by wind veer}.{\BBCQ}
\newblock
\BIn{} \APACrefbtitle {2018 Wind Energy Symposium} {2018 wind energy
  symposium}\ (\BPG~0755).
\PrintBackRefs{\CurrentBib}

\bibitem [\protect \citeauthoryear {%
{Deepwater Wind}%
}{%
{Deepwater Wind}%
}{%
{\protect \APACyear {2016}}%
}]{%
wind2016block}
\APACinsertmetastar {%
wind2016block}%
\begin{APACrefauthors}%
{Deepwater Wind}.%
\end{APACrefauthors}%
\unskip\
\newblock
\APACrefYearMonthDay{2016}{}{}.
\newblock
{\BBOQ}\APACrefatitle {Block Island Wind Farm} {Block island wind farm}.{\BBCQ}
\newblock
\APACjournalVolNumPages{URL http://dwwind.
  com/project/block-island-wind-farm}{}{}{}.
\PrintBackRefs{\CurrentBib}

\bibitem [\protect \citeauthoryear {%
Fairall%
, Markson%
, Schacher%
\BCBL {}\ \BBA {} Davidson%
}{%
Fairall%
\ \protect \BOthers {.}}{%
{\protect \APACyear {1980}}%
}]{%
fairall1980aircraft}
\APACinsertmetastar {%
fairall1980aircraft}%
\begin{APACrefauthors}%
Fairall, C.%
, Markson, R.%
, Schacher, G.%
\BCBL {}\ \BBA {} Davidson, K.%
\end{APACrefauthors}%
\unskip\
\newblock
\APACrefYearMonthDay{1980}{}{}.
\newblock
{\BBOQ}\APACrefatitle {An aircraft study of turbulence dissipation rate and
  temperature structure function in the unstable marine atmospheric boundary
  layer} {An aircraft study of turbulence dissipation rate and temperature
  structure function in the unstable marine atmospheric boundary layer}.{\BBCQ}
\newblock
\APACjournalVolNumPages{Boundary-Layer Meteorology}{19}{4}{453--469}.
\PrintBackRefs{\CurrentBib}

\bibitem [\protect \citeauthoryear {%
Filippelli%
, Markus%
, Eberhard%
, Bailey%
\BCBL {}\ \BBA {} Dubois%
}{%
Filippelli%
\ \protect \BOthers {.}}{%
{\protect \APACyear {2015}}%
}]{%
filippelli2015meocean}
\APACinsertmetastar {%
filippelli2015meocean}%
\begin{APACrefauthors}%
Filippelli, M\BPBI V.%
, Markus, M.%
, Eberhard, M.%
, Bailey, B\BPBI H.%
\BCBL {}\ \BBA {} Dubois, L.%
\end{APACrefauthors}%
\unskip\
\newblock
\APACrefYearMonthDay{2015}{}{}.
\newblock
\APACrefbtitle {Metocean Data Needs Assessment and Data Collection Strategy
  Development for the Massachusetts Wind Energy Area} {Metocean data needs
  assessment and data collection strategy development for the massachusetts
  wind energy area}\ \APACbVolEdTR{}{\BTR{}}.
\newblock
\begin{APACrefURL}
  \url{http://files.masscec.com/research/wind/MassCECMetoceanDataReport.pdf}
  \end{APACrefURL}
\PrintBackRefs{\CurrentBib}

\bibitem [\protect \citeauthoryear {%
Fleming%
\ \protect \BOthers {.}}{%
Fleming%
\ \protect \BOthers {.}}{%
{\protect \APACyear {2017}}%
}]{%
fleming2017field}
\APACinsertmetastar {%
fleming2017field}%
\begin{APACrefauthors}%
Fleming, P.%
, Annoni, J.%
, Shah, J\BPBI J.%
, Wang, L.%
, Ananthan, S.%
, Zhang, Z.%
\BDBL {}Chen, L.%
\end{APACrefauthors}%
\unskip\
\newblock
\APACrefYearMonthDay{2017}{}{}.
\newblock
{\BBOQ}\APACrefatitle {Field test of wake steering at an offshore wind farm}
  {Field test of wake steering at an offshore wind farm}.{\BBCQ}
\newblock
\APACjournalVolNumPages{Wind Energy Science}{2}{1}{229--239}.
\PrintBackRefs{\CurrentBib}

\bibitem [\protect \citeauthoryear {%
Fleming%
\ \protect \BOthers {.}}{%
Fleming%
\ \protect \BOthers {.}}{%
{\protect \APACyear {2019}}%
}]{%
fleming2019initial}
\APACinsertmetastar {%
fleming2019initial}%
\begin{APACrefauthors}%
Fleming, P.%
, King, J.%
, Dykes, K.%
, Simley, E.%
, Roadman, J.%
, Scholbrock, A.%
\BDBL {}Brake, D.%
\end{APACrefauthors}%
\unskip\
\newblock
\APACrefYearMonthDay{2019}{}{}.
\newblock
{\BBOQ}\APACrefatitle {Initial Results From a Field Campaign of Wake Steering
  Applied at a Commercial Wind Farm: Part 1} {Initial results from a field
  campaign of wake steering applied at a commercial wind farm: Part 1}.{\BBCQ}
\newblock
\APACjournalVolNumPages{Wind Energy Science Discussions}{in review}{-}{-}.
\newblock
\begin{APACrefURL} \url{https://doi.org/10.5194/wes-2019-5} \end{APACrefURL}
\PrintBackRefs{\CurrentBib}

\bibitem [\protect \citeauthoryear {%
Frehlich%
}{%
Frehlich%
}{%
{\protect \APACyear {1994}}%
}]{%
frehlich1994coherent}
\APACinsertmetastar {%
frehlich1994coherent}%
\begin{APACrefauthors}%
Frehlich, R.%
\end{APACrefauthors}%
\unskip\
\newblock
\APACrefYearMonthDay{1994}{}{}.
\newblock
{\BBOQ}\APACrefatitle {Coherent Doppler lidar signal covariance including wind
  shear and wind turbulence} {Coherent doppler lidar signal covariance
  including wind shear and wind turbulence}.{\BBCQ}
\newblock
\APACjournalVolNumPages{Applied Optics}{33}{27}{6472--6481}.
\PrintBackRefs{\CurrentBib}

\bibitem [\protect \citeauthoryear {%
Frehlich%
, Meillier%
, Jensen%
, Balsley%
\BCBL {}\ \BBA {} Sharman%
}{%
Frehlich%
\ \protect \BOthers {.}}{%
{\protect \APACyear {2006}}%
}]{%
frehlich2006measurements}
\APACinsertmetastar {%
frehlich2006measurements}%
\begin{APACrefauthors}%
Frehlich, R.%
, Meillier, Y.%
, Jensen, M\BPBI L.%
, Balsley, B.%
\BCBL {}\ \BBA {} Sharman, R.%
\end{APACrefauthors}%
\unskip\
\newblock
\APACrefYearMonthDay{2006}{}{}.
\newblock
{\BBOQ}\APACrefatitle {Measurements of boundary layer profiles in an urban
  environment} {Measurements of boundary layer profiles in an urban
  environment}.{\BBCQ}
\newblock
\APACjournalVolNumPages{Journal of Applied Meteorology and
  Climatology}{45}{6}{821--837}.
\PrintBackRefs{\CurrentBib}

\bibitem [\protect \citeauthoryear {%
Gentine%
, Pritchard%
, Rasp%
, Reinaudi%
\BCBL {}\ \BBA {} Yacalis%
}{%
Gentine%
\ \protect \BOthers {.}}{%
{\protect \APACyear {2018}}%
}]{%
gentine2018could}
\APACinsertmetastar {%
gentine2018could}%
\begin{APACrefauthors}%
Gentine, P.%
, Pritchard, M.%
, Rasp, S.%
, Reinaudi, G.%
\BCBL {}\ \BBA {} Yacalis, G.%
\end{APACrefauthors}%
\unskip\
\newblock
\APACrefYearMonthDay{2018}{}{}.
\newblock
{\BBOQ}\APACrefatitle {Could machine learning break the convection
  parameterization deadlock?} {Could machine learning break the convection
  parameterization deadlock?}{\BBCQ}
\newblock
\APACjournalVolNumPages{Geophysical Research Letters}{45}{11}{5742--5751}.
\PrintBackRefs{\CurrentBib}

\bibitem [\protect \citeauthoryear {%
Hong%
\ \BBA {} Dudhia%
}{%
Hong%
\ \BBA {} Dudhia%
}{%
{\protect \APACyear {2012}}%
}]{%
hong2012next}
\APACinsertmetastar {%
hong2012next}%
\begin{APACrefauthors}%
Hong, S\BHBI Y.%
\BCBT {}\ \BBA {} Dudhia, J.%
\end{APACrefauthors}%
\unskip\
\newblock
\APACrefYearMonthDay{2012}{}{}.
\newblock
{\BBOQ}\APACrefatitle {Next-generation numerical weather prediction: Bridging
  parameterization, explicit clouds, and large eddies} {Next-generation
  numerical weather prediction: Bridging parameterization, explicit clouds, and
  large eddies}.{\BBCQ}
\newblock
\APACjournalVolNumPages{Bulletin of the American Meteorological
  Society}{93}{1}{ES6--ES9}.
\PrintBackRefs{\CurrentBib}

\bibitem [\protect \citeauthoryear {%
Krishnamurthy%
, Calhoun%
, Billings%
\BCBL {}\ \BBA {} Doyle%
}{%
Krishnamurthy%
\ \protect \BOthers {.}}{%
{\protect \APACyear {2011}}%
}]{%
krishnamurthy2011wind}
\APACinsertmetastar {%
krishnamurthy2011wind}%
\begin{APACrefauthors}%
Krishnamurthy, R.%
, Calhoun, R.%
, Billings, B.%
\BCBL {}\ \BBA {} Doyle, J.%
\end{APACrefauthors}%
\unskip\
\newblock
\APACrefYearMonthDay{2011}{}{}.
\newblock
{\BBOQ}\APACrefatitle {Wind turbulence estimates in a valley by coherent
  Doppler lidar} {Wind turbulence estimates in a valley by coherent doppler
  lidar}.{\BBCQ}
\newblock
\APACjournalVolNumPages{Meteorological Applications}{18}{3}{361--371}.
\PrintBackRefs{\CurrentBib}

\bibitem [\protect \citeauthoryear {%
Kumer%
, Reuder%
, Dorninger%
, Zauner%
\BCBL {}\ \BBA {} Grubi{\v{s}}i{\'c}%
}{%
Kumer%
\ \protect \BOthers {.}}{%
{\protect \APACyear {2016}}%
}]{%
kumer2016turbulent}
\APACinsertmetastar {%
kumer2016turbulent}%
\begin{APACrefauthors}%
Kumer, V\BHBI M.%
, Reuder, J.%
, Dorninger, M.%
, Zauner, R.%
\BCBL {}\ \BBA {} Grubi{\v{s}}i{\'c}, V.%
\end{APACrefauthors}%
\unskip\
\newblock
\APACrefYearMonthDay{2016}{}{}.
\newblock
{\BBOQ}\APACrefatitle {Turbulent kinetic energy estimates from profiling wind
  LiDAR measurements and their potential for wind energy applications}
  {Turbulent kinetic energy estimates from profiling wind lidar measurements
  and their potential for wind energy applications}.{\BBCQ}
\newblock
\APACjournalVolNumPages{Renewable Energy}{99}{}{898--910}.
\PrintBackRefs{\CurrentBib}

\bibitem [\protect \citeauthoryear {%
Landberg%
}{%
Landberg%
}{%
{\protect \APACyear {2015}}%
}]{%
landberg2015meteorology}
\APACinsertmetastar {%
landberg2015meteorology}%
\begin{APACrefauthors}%
Landberg, L.%
\end{APACrefauthors}%
\unskip\
\newblock
\APACrefYear{2015}.
\newblock
\APACrefbtitle {Meteorology for Wind Energy: An Introduction} {Meteorology for
  wind energy: An introduction}.
\newblock
\APACaddressPublisher{Hoboken, New Jersey}{John Wiley \& Sons}.
\PrintBackRefs{\CurrentBib}

\bibitem [\protect \citeauthoryear {%
Lawrence%
\ \BBA {} Balsley%
}{%
Lawrence%
\ \BBA {} Balsley%
}{%
{\protect \APACyear {2013}}%
}]{%
lawrence2013high}
\APACinsertmetastar {%
lawrence2013high}%
\begin{APACrefauthors}%
Lawrence, D\BPBI A.%
\BCBT {}\ \BBA {} Balsley, B\BPBI B.%
\end{APACrefauthors}%
\unskip\
\newblock
\APACrefYearMonthDay{2013}{}{}.
\newblock
{\BBOQ}\APACrefatitle {High-resolution atmospheric sensing of multiple
  atmospheric variables using the DataHawk small airborne measurement system}
  {High-resolution atmospheric sensing of multiple atmospheric variables using
  the datahawk small airborne measurement system}.{\BBCQ}
\newblock
\APACjournalVolNumPages{Journal of Atmospheric and Oceanic
  Technology}{30}{10}{2352--2366}.
\PrintBackRefs{\CurrentBib}

\bibitem [\protect \citeauthoryear {%
Lundquist%
\ \BBA {} Bariteau%
}{%
Lundquist%
\ \BBA {} Bariteau%
}{%
{\protect \APACyear {2015}}%
}]{%
lundquist2015dissipation}
\APACinsertmetastar {%
lundquist2015dissipation}%
\begin{APACrefauthors}%
Lundquist, J\BPBI K.%
\BCBT {}\ \BBA {} Bariteau, L.%
\end{APACrefauthors}%
\unskip\
\newblock
\APACrefYearMonthDay{2015}{}{}.
\newblock
{\BBOQ}\APACrefatitle {Dissipation of Turbulence in the Wake of a Wind Turbine}
  {Dissipation of turbulence in the wake of a wind turbine}.{\BBCQ}
\newblock
\APACjournalVolNumPages{Boundary-Layer Meteorology}{154}{2}{229--241}.
\PrintBackRefs{\CurrentBib}

\bibitem [\protect \citeauthoryear {%
Lundquist%
, DuVivier%
, Kaffine%
\BCBL {}\ \BBA {} Tomaszewski%
}{%
Lundquist%
\ \protect \BOthers {.}}{%
{\protect \APACyear {2019}}%
}]{%
lundquist2019costs}
\APACinsertmetastar {%
lundquist2019costs}%
\begin{APACrefauthors}%
Lundquist, J\BPBI K.%
, DuVivier, K\BPBI K.%
, Kaffine, D.%
\BCBL {}\ \BBA {} Tomaszewski, J\BPBI M.%
\end{APACrefauthors}%
\unskip\
\newblock
\APACrefYearMonthDay{2019}{}{}.
\newblock
{\BBOQ}\APACrefatitle {Costs and consequences of wind turbine wake effects
  arising from uncoordinated wind energy development} {Costs and consequences
  of wind turbine wake effects arising from uncoordinated wind energy
  development}.{\BBCQ}
\newblock
\APACjournalVolNumPages{Nature Energy}{4}{1}{26--34}.
\PrintBackRefs{\CurrentBib}

\bibitem [\protect \citeauthoryear {%
Macknick%
, Newmark%
, Heath%
\BCBL {}\ \BBA {} Hallett%
}{%
Macknick%
\ \protect \BOthers {.}}{%
{\protect \APACyear {2012}}%
}]{%
macknick2012operational}
\APACinsertmetastar {%
macknick2012operational}%
\begin{APACrefauthors}%
Macknick, J.%
, Newmark, R.%
, Heath, G.%
\BCBL {}\ \BBA {} Hallett, K\BPBI C.%
\end{APACrefauthors}%
\unskip\
\newblock
\APACrefYearMonthDay{2012}{}{}.
\newblock
{\BBOQ}\APACrefatitle {Operational water consumption and withdrawal factors for
  electricity generating technologies: a review of existing literature}
  {Operational water consumption and withdrawal factors for electricity
  generating technologies: a review of existing literature}.{\BBCQ}
\newblock
\APACjournalVolNumPages{Environmental Research Letters}{7}{4}{045802}.
\PrintBackRefs{\CurrentBib}

\bibitem [\protect \citeauthoryear {%
Manwell%
, McGowan%
\BCBL {}\ \BBA {} Rogers%
}{%
Manwell%
\ \protect \BOthers {.}}{%
{\protect \APACyear {2010}}%
}]{%
manwell2010wind}
\APACinsertmetastar {%
manwell2010wind}%
\begin{APACrefauthors}%
Manwell, J\BPBI F.%
, McGowan, J\BPBI G.%
\BCBL {}\ \BBA {} Rogers, A\BPBI L.%
\end{APACrefauthors}%
\unskip\
\newblock
\APACrefYear{2010}.
\newblock
\APACrefbtitle {Wind energy explained: theory, design and application} {Wind
  energy explained: theory, design and application}.
\newblock
\APACaddressPublisher{Hoboken, New Jersey}{John Wiley \& Sons}.
\PrintBackRefs{\CurrentBib}

\bibitem [\protect \citeauthoryear {%
{Massachusetts Clean Energy Center}%
, {Bristol Community College}%
, {UMass DartMouth Public Policy Center}%
\BCBL {}\ \BBA {} {Massachusetts Maritime Academy}%
}{%
{Massachusetts Clean Energy Center}%
\ \protect \BOthers {.}}{%
{\protect \APACyear {2018}}%
}]{%
center20182018}
\APACinsertmetastar {%
center20182018}%
\begin{APACrefauthors}%
{Massachusetts Clean Energy Center}%
, {Bristol Community College}%
, {UMass DartMouth Public Policy Center}%
\BCBL {}\ \BBA {} {Massachusetts Maritime Academy}.%
\end{APACrefauthors}%
\unskip\
\newblock
\APACrefYear{2018}.
\newblock
\APACrefbtitle {2018 Massachusetts Offshore Wind Workforce Assessment} {2018
  massachusetts offshore wind workforce assessment}.
\newblock
\APACaddressPublisher{}{Massachusetts Clean Energy Center}.
\PrintBackRefs{\CurrentBib}

\bibitem [\protect \citeauthoryear {%
McCaffrey%
, Bianco%
\BCBL {}\ \BBA {} Wilczak%
}{%
McCaffrey%
\ \protect \BOthers {.}}{%
{\protect \APACyear {2017}}%
}]{%
mccaffrey2017improved}
\APACinsertmetastar {%
mccaffrey2017improved}%
\begin{APACrefauthors}%
McCaffrey, K.%
, Bianco, L.%
\BCBL {}\ \BBA {} Wilczak, J\BPBI M.%
\end{APACrefauthors}%
\unskip\
\newblock
\APACrefYearMonthDay{2017}{}{}.
\newblock
{\BBOQ}\APACrefatitle {Improved observations of turbulence dissipation rates
  from wind profiling radars} {Improved observations of turbulence dissipation
  rates from wind profiling radars}.{\BBCQ}
\newblock
\APACjournalVolNumPages{Atmospheric Measurement Techniques}{10}{7}{2595--2611}.
\newblock
\begin{APACrefDOI} \doi{10.5194/amt-10-2595-2017} \end{APACrefDOI}
\PrintBackRefs{\CurrentBib}

\bibitem [\protect \citeauthoryear {%
Musial%
\ \protect \BOthers {.}}{%
Musial%
\ \protect \BOthers {.}}{%
{\protect \APACyear {2017}}%
}]{%
musial20172016}
\APACinsertmetastar {%
musial20172016}%
\begin{APACrefauthors}%
Musial, W.%
, Beiter, P.%
, Schwabe, P.%
, Tian, T.%
, Stehly, T.%
, Spitsen, P.%
\BDBL {}Gevorgian, V.%
\end{APACrefauthors}%
\unskip\
\newblock
\APACrefYearMonthDay{2017}{}{}.
\newblock
\APACrefbtitle {2016 Offshore Wind Technologies Market Report} {2016 offshore
  wind technologies market report}\ \APACbVolEdTR{}{\BTR{}}.
\newblock
\APACaddressInstitution{}{National Renewable Energy Laboratory (NREL), Golden,
  CO (United States)}.
\newblock
\begin{APACrefURL}
  \url{https://www.energy.gov/sites/prod/files/2017/08/f35/2016%20Offshore%20Wind%20Technologies%20Market%20Report.pdf}
  \end{APACrefURL}
\PrintBackRefs{\CurrentBib}

\bibitem [\protect \citeauthoryear {%
Musial%
, Heimiller%
, Beiter%
, Scott%
\BCBL {}\ \BBA {} Draxl%
}{%
Musial%
\ \protect \BOthers {.}}{%
{\protect \APACyear {2016}}%
}]{%
musial2016offshore}
\APACinsertmetastar {%
musial2016offshore}%
\begin{APACrefauthors}%
Musial, W.%
, Heimiller, D.%
, Beiter, P.%
, Scott, G.%
\BCBL {}\ \BBA {} Draxl, C.%
\end{APACrefauthors}%
\unskip\
\newblock
\APACrefYearMonthDay{2016}{}{}.
\newblock
\APACrefbtitle {Offshore wind energy resource assessment for the United States}
  {Offshore wind energy resource assessment for the united states}\
  \APACbVolEdTR{}{\BTR{}}.
\newblock
\APACaddressInstitution{}{National Renewable Energy Laboratory (NREL), Golden,
  CO (United States)}.
\newblock
\begin{APACrefURL} \url{https://www.nrel.gov/docs/fy16osti/66599.pdf}
  \end{APACrefURL}
\PrintBackRefs{\CurrentBib}

\bibitem [\protect \citeauthoryear {%
Nakanishi%
\ \BBA {} Niino%
}{%
Nakanishi%
\ \BBA {} Niino%
}{%
{\protect \APACyear {2006}}%
}]{%
nakanishi2006improved}
\APACinsertmetastar {%
nakanishi2006improved}%
\begin{APACrefauthors}%
Nakanishi, M.%
\BCBT {}\ \BBA {} Niino, H.%
\end{APACrefauthors}%
\unskip\
\newblock
\APACrefYearMonthDay{2006}{}{}.
\newblock
{\BBOQ}\APACrefatitle {An improved {M}ellor--{Y}amada level-3 model: Its
  numerical stability and application to a regional prediction of advection
  fog} {An improved {M}ellor--{Y}amada level-3 model: Its numerical stability
  and application to a regional prediction of advection fog}.{\BBCQ}
\newblock
\APACjournalVolNumPages{Boundary-Layer Meteorology}{119}{2}{397--407}.
\PrintBackRefs{\CurrentBib}

\bibitem [\protect \citeauthoryear {%
Nygaard%
}{%
Nygaard%
}{%
{\protect \APACyear {2014}}%
}]{%
nygaard2014wakes}
\APACinsertmetastar {%
nygaard2014wakes}%
\begin{APACrefauthors}%
Nygaard, N\BPBI G.%
\end{APACrefauthors}%
\unskip\
\newblock
\APACrefYearMonthDay{2014}{}{}.
\newblock
{\BBOQ}\APACrefatitle {Wakes in very large wind farms and the effect of
  neighbouring wind farms} {Wakes in very large wind farms and the effect of
  neighbouring wind farms}.{\BBCQ}
\newblock
\BIn{} \APACrefbtitle {Journal of Physics: Conference Series} {Journal of
  physics: Conference series}\ (\BVOL~524, \BPG~012162).
\PrintBackRefs{\CurrentBib}

\bibitem [\protect \citeauthoryear {%
O'Connor%
\ \protect \BOthers {.}}{%
O'Connor%
\ \protect \BOthers {.}}{%
{\protect \APACyear {2010}}%
}]{%
o2010method}
\APACinsertmetastar {%
o2010method}%
\begin{APACrefauthors}%
O'Connor, E\BPBI J.%
, Illingworth, A\BPBI J.%
, Brooks, I\BPBI M.%
, Westbrook, C\BPBI D.%
, Hogan, R\BPBI J.%
, Davies, F.%
\BCBL {}\ \BBA {} Brooks, B\BPBI J.%
\end{APACrefauthors}%
\unskip\
\newblock
\APACrefYearMonthDay{2010}{}{}.
\newblock
{\BBOQ}\APACrefatitle {A method for estimating the turbulent kinetic energy
  dissipation rate from a vertically pointing Doppler lidar, and independent
  evaluation from balloon-borne in situ measurements} {A method for estimating
  the turbulent kinetic energy dissipation rate from a vertically pointing
  doppler lidar, and independent evaluation from balloon-borne in situ
  measurements}.{\BBCQ}
\newblock
\APACjournalVolNumPages{Journal of Atmospheric and Oceanic
  Technology}{27}{10}{1652--1664}.
\PrintBackRefs{\CurrentBib}

\bibitem [\protect \citeauthoryear {%
Oncley%
\ \protect \BOthers {.}}{%
Oncley%
\ \protect \BOthers {.}}{%
{\protect \APACyear {1996}}%
}]{%
oncley1996surface}
\APACinsertmetastar {%
oncley1996surface}%
\begin{APACrefauthors}%
Oncley, S\BPBI P.%
, Friehe, C\BPBI A.%
, Larue, J\BPBI C.%
, Businger, J\BPBI A.%
, Itsweire, E\BPBI C.%
\BCBL {}\ \BBA {} Chang, S\BPBI S.%
\end{APACrefauthors}%
\unskip\
\newblock
\APACrefYearMonthDay{1996}{}{}.
\newblock
{\BBOQ}\APACrefatitle {Surface-layer fluxes, profiles, and turbulence
  measurements over uniform terrain under near-neutral conditions}
  {Surface-layer fluxes, profiles, and turbulence measurements over uniform
  terrain under near-neutral conditions}.{\BBCQ}
\newblock
\APACjournalVolNumPages{Journal of the Atmospheric
  Sciences}{53}{7}{1029--1044}.
\PrintBackRefs{\CurrentBib}

\bibitem [\protect \citeauthoryear {%
Paquin%
\ \BBA {} Pond%
}{%
Paquin%
\ \BBA {} Pond%
}{%
{\protect \APACyear {1971}}%
}]{%
paquin1971determination}
\APACinsertmetastar {%
paquin1971determination}%
\begin{APACrefauthors}%
Paquin, J.%
\BCBT {}\ \BBA {} Pond, S.%
\end{APACrefauthors}%
\unskip\
\newblock
\APACrefYearMonthDay{1971}{}{}.
\newblock
{\BBOQ}\APACrefatitle {The determination of the Kolmogoroff constants for
  velocity, temperature and humidity fluctuations from second-and third-order
  structure functions} {The determination of the kolmogoroff constants for
  velocity, temperature and humidity fluctuations from second-and third-order
  structure functions}.{\BBCQ}
\newblock
\APACjournalVolNumPages{Journal of Fluid Mechanics}{50}{2}{257--269}.
\PrintBackRefs{\CurrentBib}

\bibitem [\protect \citeauthoryear {%
Pearson%
, Davies%
\BCBL {}\ \BBA {} Collier%
}{%
Pearson%
\ \protect \BOthers {.}}{%
{\protect \APACyear {2009}}%
}]{%
pearson2009analysis}
\APACinsertmetastar {%
pearson2009analysis}%
\begin{APACrefauthors}%
Pearson, G.%
, Davies, F.%
\BCBL {}\ \BBA {} Collier, C.%
\end{APACrefauthors}%
\unskip\
\newblock
\APACrefYearMonthDay{2009}{}{}.
\newblock
{\BBOQ}\APACrefatitle {An analysis of the performance of the UFAM pulsed
  Doppler lidar for observing the boundary layer} {An analysis of the
  performance of the ufam pulsed doppler lidar for observing the boundary
  layer}.{\BBCQ}
\newblock
\APACjournalVolNumPages{Journal of Atmospheric and Oceanic
  Technology}{26}{2}{240--250}.
\PrintBackRefs{\CurrentBib}

\bibitem [\protect \citeauthoryear {%
Platis%
\ \protect \BOthers {.}}{%
Platis%
\ \protect \BOthers {.}}{%
{\protect \APACyear {2018}}%
}]{%
platis2018first}
\APACinsertmetastar {%
platis2018first}%
\begin{APACrefauthors}%
Platis, A.%
, Siedersleben, S\BPBI K.%
, Bange, J.%
, Lampert, A.%
, B{\"a}rfuss, K.%
, Hankers, R.%
\BDBL {}others%
\end{APACrefauthors}%
\unskip\
\newblock
\APACrefYearMonthDay{2018}{}{}.
\newblock
{\BBOQ}\APACrefatitle {First in situ evidence of wakes in the far field behind
  offshore wind farms} {First in situ evidence of wakes in the far field behind
  offshore wind farms}.{\BBCQ}
\newblock
\APACjournalVolNumPages{Scientific reports}{8}{1}{2163}.
\PrintBackRefs{\CurrentBib}

\bibitem [\protect \citeauthoryear {%
Rhodes%
\ \BBA {} Lundquist%
}{%
Rhodes%
\ \BBA {} Lundquist%
}{%
{\protect \APACyear {2013}}%
}]{%
rhodes_effect_2013}
\APACinsertmetastar {%
rhodes_effect_2013}%
\begin{APACrefauthors}%
Rhodes, M\BPBI E.%
\BCBT {}\ \BBA {} Lundquist, J\BPBI K.%
\end{APACrefauthors}%
\unskip\
\newblock
\APACrefYearMonthDay{2013}{{\APACmonth{10}}}{}.
\newblock
{\BBOQ}\APACrefatitle {The {Effect} of {Wind}-{Turbine} {Wakes} on {Summertime}
  {US} {Midwest} {Atmospheric} {Wind} {Profiles} as {Observed} with
  {Ground}-{Based} {Doppler} {Lidar}} {The {Effect} of {Wind}-{Turbine} {Wakes}
  on {Summertime} {US} {Midwest} {Atmospheric} {Wind} {Profiles} as {Observed}
  with {Ground}-{Based} {Doppler} {Lidar}}.{\BBCQ}
\newblock
\APACjournalVolNumPages{Boundary-Layer Meteorology}{149}{1}{85--103}.
\newblock
\begin{APACrefDOI} \doi{10.1007/s10546-013-9834-x} \end{APACrefDOI}
\PrintBackRefs{\CurrentBib}

\bibitem [\protect \citeauthoryear {%
Sathe%
, Mann%
, Gottschall%
\BCBL {}\ \BBA {} Courtney%
}{%
Sathe%
\ \protect \BOthers {.}}{%
{\protect \APACyear {2011}}%
}]{%
sathe2011can}
\APACinsertmetastar {%
sathe2011can}%
\begin{APACrefauthors}%
Sathe, A.%
, Mann, J.%
, Gottschall, J.%
\BCBL {}\ \BBA {} Courtney, M.%
\end{APACrefauthors}%
\unskip\
\newblock
\APACrefYearMonthDay{2011}{}{}.
\newblock
{\BBOQ}\APACrefatitle {Can wind lidars measure turbulence?} {Can wind lidars
  measure turbulence?}{\BBCQ}
\newblock
\APACjournalVolNumPages{Journal of Atmospheric and Oceanic
  Technology}{28}{7}{853--868}.
\PrintBackRefs{\CurrentBib}

\bibitem [\protect \citeauthoryear {%
Sharma%
, Sharma%
, Irwin%
\BCBL {}\ \BBA {} Shenoy%
}{%
Sharma%
\ \protect \BOthers {.}}{%
{\protect \APACyear {2011}}%
}]{%
sharma2011predicting}
\APACinsertmetastar {%
sharma2011predicting}%
\begin{APACrefauthors}%
Sharma, N.%
, Sharma, P.%
, Irwin, D.%
\BCBL {}\ \BBA {} Shenoy, P.%
\end{APACrefauthors}%
\unskip\
\newblock
\APACrefYearMonthDay{2011}{}{}.
\newblock
{\BBOQ}\APACrefatitle {Predicting solar generation from weather forecasts using
  machine learning} {Predicting solar generation from weather forecasts using
  machine learning}.{\BBCQ}
\newblock
\BIn{} \APACrefbtitle {2011 {IEEE} {I}nternational {C}onference on {S}mart
  {G}rid {C}ommunications} {2011 {IEEE} {I}nternational {C}onference on {S}mart
  {G}rid {C}ommunications}\ (\BPGS\ 528--533).
\PrintBackRefs{\CurrentBib}

\bibitem [\protect \citeauthoryear {%
Shaw%
\ \BBA {} LeMone%
}{%
Shaw%
\ \BBA {} LeMone%
}{%
{\protect \APACyear {2003}}%
}]{%
shaw2003turbulence}
\APACinsertmetastar {%
shaw2003turbulence}%
\begin{APACrefauthors}%
Shaw, W\BPBI J.%
\BCBT {}\ \BBA {} LeMone, M\BPBI A.%
\end{APACrefauthors}%
\unskip\
\newblock
\APACrefYearMonthDay{2003}{}{}.
\newblock
{\BBOQ}\APACrefatitle {Turbulence dissipation rate measured by 915 MHz wind
  profiling radars compared with in-situ tower and aircraft data} {Turbulence
  dissipation rate measured by 915 mhz wind profiling radars compared with
  in-situ tower and aircraft data}.{\BBCQ}
\newblock
\BIn{} \APACrefbtitle {12th Symposium on Meteorological Observations and
  Instrumentation.} {12th symposium on meteorological observations and
  instrumentation.}
\newblock
\begin{APACrefURL} \url{https://ams.confex.com/ams/pdfpapers/58647.pdf}
  \end{APACrefURL}
\PrintBackRefs{\CurrentBib}

\bibitem [\protect \citeauthoryear {%
Siedersleben%
, Lundquist%
\BCBL {}\ \protect \BOthers {.}}{%
Siedersleben%
, Lundquist%
\BCBL {}\ \protect \BOthers {.}}{%
{\protect \APACyear {2018}}%
}]{%
siedersleben2018micrometeorological}
\APACinsertmetastar {%
siedersleben2018micrometeorological}%
\begin{APACrefauthors}%
Siedersleben, S\BPBI K.%
, Lundquist, J\BPBI K.%
, Platis, A.%
, Bange, J.%
, B{\"a}rfuss, K.%
, Lampert, A.%
\BDBL {}Emeis, S.%
\end{APACrefauthors}%
\unskip\
\newblock
\APACrefYearMonthDay{2018}{}{}.
\newblock
{\BBOQ}\APACrefatitle {Micrometeorological impacts of offshore wind farms as
  seen in observations and simulations} {Micrometeorological impacts of
  offshore wind farms as seen in observations and simulations}.{\BBCQ}
\newblock
\APACjournalVolNumPages{Environmental Research Letters}{13}{12}{124012}.
\PrintBackRefs{\CurrentBib}

\bibitem [\protect \citeauthoryear {%
Siedersleben%
, Platis%
\BCBL {}\ \protect \BOthers {.}}{%
Siedersleben%
, Platis%
\BCBL {}\ \protect \BOthers {.}}{%
{\protect \APACyear {2018}}%
}]{%
siedersleben2018evaluation}
\APACinsertmetastar {%
siedersleben2018evaluation}%
\begin{APACrefauthors}%
Siedersleben, S\BPBI K.%
, Platis, A.%
, Lundquist, J\BPBI K.%
, Lampert, A.%
, B{\"a}rfuss, K.%
, Ca{\~n}adillas, B.%
\BDBL {}others%
\end{APACrefauthors}%
\unskip\
\newblock
\APACrefYearMonthDay{2018}{}{}.
\newblock
{\BBOQ}\APACrefatitle {Evaluation of a wind farm parametrization for mesoscale
  atmospheric flow models with aircraft measurements} {Evaluation of a wind
  farm parametrization for mesoscale atmospheric flow models with aircraft
  measurements}.{\BBCQ}
\newblock
\APACjournalVolNumPages{Meteorologische Zeitschrift}{}{}{}.
\PrintBackRefs{\CurrentBib}

\bibitem [\protect \citeauthoryear {%
Skamarock%
\ \protect \BOthers {.}}{%
Skamarock%
\ \protect \BOthers {.}}{%
{\protect \APACyear {2005}}%
}]{%
skamarock2005description}
\APACinsertmetastar {%
skamarock2005description}%
\begin{APACrefauthors}%
Skamarock, W\BPBI C.%
, Klemp, J\BPBI B.%
, Dudhia, J.%
, Gill, D\BPBI O.%
, Barker, D\BPBI M.%
, Wang, W.%
\BCBL {}\ \BBA {} Powers, J\BPBI G.%
\end{APACrefauthors}%
\unskip\
\newblock
\APACrefYearMonthDay{2005}{}{}.
\newblock
\APACrefbtitle {A description of the advanced research WRF version 2} {A
  description of the advanced research wrf version 2}\ \APACbVolEdTR{}{\BTR{}}.
\newblock
\APACaddressInstitution{}{National Center For Atmospheric Research Boulder Co
  Mesoscale and Microscale Meteorology Div}.
\PrintBackRefs{\CurrentBib}

\bibitem [\protect \citeauthoryear {%
Sreenivasan%
}{%
Sreenivasan%
}{%
{\protect \APACyear {1995}}%
}]{%
sreenivasan1995universality}
\APACinsertmetastar {%
sreenivasan1995universality}%
\begin{APACrefauthors}%
Sreenivasan, K\BPBI R.%
\end{APACrefauthors}%
\unskip\
\newblock
\APACrefYearMonthDay{1995}{}{}.
\newblock
{\BBOQ}\APACrefatitle {On the universality of the Kolmogorov constant} {On the
  universality of the kolmogorov constant}.{\BBCQ}
\newblock
\APACjournalVolNumPages{Physics of Fluids}{7}{11}{2778--2784}.
\PrintBackRefs{\CurrentBib}

\bibitem [\protect \citeauthoryear {%
Stiesdal%
}{%
Stiesdal%
}{%
{\protect \APACyear {2016}}%
}]{%
stiesdal_2016}
\APACinsertmetastar {%
stiesdal_2016}%
\begin{APACrefauthors}%
Stiesdal, H.%
\end{APACrefauthors}%
\unskip\
\newblock
\APACrefYearMonthDay{2016}{Dec}{}.
\newblock
{\BBOQ}\APACrefatitle {Midt i en disruptionstid} {Midt i en
  disruptionstid}.{\BBCQ}
\newblock
\APACjournalVolNumPages{Ingeniøren}{}{}{}.
\newblock
\begin{APACrefURL} \url{https://ing.dk/blog/midt-disruptionstid-190449}
  \end{APACrefURL}
\PrintBackRefs{\CurrentBib}

\bibitem [\protect \citeauthoryear {%
Tonttila%
\ \protect \BOthers {.}}{%
Tonttila%
\ \protect \BOthers {.}}{%
{\protect \APACyear {2015}}%
}]{%
tonttila2015turbulent}
\APACinsertmetastar {%
tonttila2015turbulent}%
\begin{APACrefauthors}%
Tonttila, J.%
, O'Connor, E.%
, Hellsten, A.%
, Hirsikko, A.%
, O'Dowd, C.%
, J{\"a}rvinen, H.%
\BCBL {}\ \BBA {} R{\"a}is{\"a}nen, P.%
\end{APACrefauthors}%
\unskip\
\newblock
\APACrefYearMonthDay{2015}{}{}.
\newblock
{\BBOQ}\APACrefatitle {Turbulent structure and scaling of the inertial subrange
  in a stratocumulus-topped boundary layer observed by a Doppler lidar}
  {Turbulent structure and scaling of the inertial subrange in a
  stratocumulus-topped boundary layer observed by a doppler lidar}.{\BBCQ}
\newblock
\APACjournalVolNumPages{Atmospheric Chemistry and Physics}{15}{10}{5873--5885}.
\PrintBackRefs{\CurrentBib}

\bibitem [\protect \citeauthoryear {%
van Hoof%
}{%
van Hoof%
}{%
{\protect \APACyear {2017}}%
}]{%
hoof2017unlocking}
\APACinsertmetastar {%
hoof2017unlocking}%
\begin{APACrefauthors}%
van Hoof, J.%
\end{APACrefauthors}%
\unskip\
\newblock
\APACrefYearMonthDay{2017}{}{}.
\newblock
\APACrefbtitle {Unlocking Europe’s offshore wind potential} {Unlocking
  europe’s offshore wind potential}\ \APACbVolEdTR{}{\BTR{}}.
\newblock
\APACaddressInstitution{}{PricewaterhouseCoopers B.V.}
\newblock
\begin{APACrefURL}
  \url{https://www.pwc.nl/nl/assets/documents/pwc-unlocking-europes-offshore-wind-potential.pdf}
  \end{APACrefURL}
\PrintBackRefs{\CurrentBib}

\bibitem [\protect \citeauthoryear {%
Walter%
, Weiss%
, Swift%
, Chapman%
\BCBL {}\ \BBA {} Kelley%
}{%
Walter%
\ \protect \BOthers {.}}{%
{\protect \APACyear {2009}}%
}]{%
walter2009speed}
\APACinsertmetastar {%
walter2009speed}%
\begin{APACrefauthors}%
Walter, K.%
, Weiss, C\BPBI C.%
, Swift, A\BPBI H.%
, Chapman, J.%
\BCBL {}\ \BBA {} Kelley, N\BPBI D.%
\end{APACrefauthors}%
\unskip\
\newblock
\APACrefYearMonthDay{2009}{}{}.
\newblock
{\BBOQ}\APACrefatitle {Speed and direction shear in the stable nocturnal
  boundary layer} {Speed and direction shear in the stable nocturnal boundary
  layer}.{\BBCQ}
\newblock
\APACjournalVolNumPages{Journal of Solar Energy Engineering}{131}{1}{011013}.
\PrintBackRefs{\CurrentBib}

\bibitem [\protect \citeauthoryear {%
Wiser%
\ \protect \BOthers {.}}{%
Wiser%
\ \protect \BOthers {.}}{%
{\protect \APACyear {2015}}%
}]{%
wiser2015wind}
\APACinsertmetastar {%
wiser2015wind}%
\begin{APACrefauthors}%
Wiser, R.%
, Lantz, E.%
, Mai, T.%
, Zayas, J.%
, DeMeo, E.%
, Eugeni, E.%
\BDBL {}Tusing, R.%
\end{APACrefauthors}%
\unskip\
\newblock
\APACrefYearMonthDay{2015}{}{}.
\newblock
{\BBOQ}\APACrefatitle {Wind vision: A new era for wind power in the United
  States} {Wind vision: A new era for wind power in the united states}.{\BBCQ}
\newblock
\APACjournalVolNumPages{The Electricity Journal}{28}{9}{120--132}.
\PrintBackRefs{\CurrentBib}

\bibitem [\protect \citeauthoryear {%
Worsnop%
, Lundquist%
, Bryan%
, Damiani%
\BCBL {}\ \BBA {} Musial%
}{%
Worsnop%
\ \protect \BOthers {.}}{%
{\protect \APACyear {2017}}%
}]{%
worsnop2017gusts}
\APACinsertmetastar {%
worsnop2017gusts}%
\begin{APACrefauthors}%
Worsnop, R\BPBI P.%
, Lundquist, J\BPBI K.%
, Bryan, G\BPBI H.%
, Damiani, R.%
\BCBL {}\ \BBA {} Musial, W.%
\end{APACrefauthors}%
\unskip\
\newblock
\APACrefYearMonthDay{2017}{}{}.
\newblock
{\BBOQ}\APACrefatitle {Gusts and shear within hurricane eyewalls can exceed
  offshore wind turbine design standards} {Gusts and shear within hurricane
  eyewalls can exceed offshore wind turbine design standards}.{\BBCQ}
\newblock
\APACjournalVolNumPages{Geophysical Research Letters}{44}{12}{6413--6420}.
\PrintBackRefs{\CurrentBib}

\bibitem [\protect \citeauthoryear {%
Wulfmeyer%
\ \protect \BOthers {.}}{%
Wulfmeyer%
\ \protect \BOthers {.}}{%
{\protect \APACyear {2016}}%
}]{%
wulfmeyer2016determination}
\APACinsertmetastar {%
wulfmeyer2016determination}%
\begin{APACrefauthors}%
Wulfmeyer, V.%
, Muppa, S\BPBI K.%
, Behrendt, A.%
, Hammann, E.%
, Sp{\"a}th, F.%
, Sorbjan, Z.%
\BDBL {}Hardesty, R\BPBI M.%
\end{APACrefauthors}%
\unskip\
\newblock
\APACrefYearMonthDay{2016}{}{}.
\newblock
{\BBOQ}\APACrefatitle {Determination of convective boundary layer entrainment
  fluxes, dissipation rates, and the molecular destruction of variances:
  Theoretical description and a strategy for its confirmation with a novel
  lidar system synergy} {Determination of convective boundary layer entrainment
  fluxes, dissipation rates, and the molecular destruction of variances:
  Theoretical description and a strategy for its confirmation with a novel
  lidar system synergy}.{\BBCQ}
\newblock
\APACjournalVolNumPages{Journal of the Atmospheric Sciences}{73}{2}{667--692}.
\PrintBackRefs{\CurrentBib}

\bibitem [\protect \citeauthoryear {%
Xingjian%
\ \protect \BOthers {.}}{%
Xingjian%
\ \protect \BOthers {.}}{%
{\protect \APACyear {2015}}%
}]{%
xingjian2015convolutional}
\APACinsertmetastar {%
xingjian2015convolutional}%
\begin{APACrefauthors}%
Xingjian, S.%
, Chen, Z.%
, Wang, H.%
, Yeung, D\BHBI Y.%
, Wong, W\BHBI K.%
\BCBL {}\ \BBA {} Woo, W\BHBI C.%
\end{APACrefauthors}%
\unskip\
\newblock
\APACrefYearMonthDay{2015}{}{}.
\newblock
{\BBOQ}\APACrefatitle {Convolutional {LSTM} network: A machine learning
  approach for precipitation nowcasting} {Convolutional {LSTM} network: A
  machine learning approach for precipitation nowcasting}.{\BBCQ}
\newblock
\BIn{} \APACrefbtitle {Advances in Neural Information Processing Systems}
  {Advances in neural information processing systems}\ (\BPGS\ 802--810).
\PrintBackRefs{\CurrentBib}

\bibitem [\protect \citeauthoryear {%
Yang%
\ \protect \BOthers {.}}{%
Yang%
\ \protect \BOthers {.}}{%
{\protect \APACyear {2017}}%
}]{%
yang2017sensitivity}
\APACinsertmetastar {%
yang2017sensitivity}%
\begin{APACrefauthors}%
Yang, B.%
, Qian, Y.%
, Berg, L\BPBI K.%
, Ma, P\BHBI L.%
, Wharton, S.%
, Bulaevskaya, V.%
\BDBL {}Shaw, W\BPBI J.%
\end{APACrefauthors}%
\unskip\
\newblock
\APACrefYearMonthDay{2017}{}{}.
\newblock
{\BBOQ}\APACrefatitle {Sensitivity of turbine-height wind speeds to parameters
  in planetary boundary-layer and surface-layer schemes in the weather research
  and forecasting model} {Sensitivity of turbine-height wind speeds to
  parameters in planetary boundary-layer and surface-layer schemes in the
  weather research and forecasting model}.{\BBCQ}
\newblock
\APACjournalVolNumPages{Boundary-Layer Meteorology}{162}{1}{117--142}.
\PrintBackRefs{\CurrentBib}

\end{thebibliography}
	
\end{document}